\documentclass[10pt,bibnotes,notitlepage,final,balancelastpage,superscriptaddress,twocolumn,showpacs,pra]{revtex4}
\usepackage{graphicx}
\usepackage{amssymb, amsmath, amsfonts, color, rotating, multirow, graphicx, bm}
\usepackage[dvipdfm,bookmarks=false,pdfstartview=FitH,hyperindex=true, colorlinks, linkcolor=blue, citecolor=blue]{hyperref}
\begin{document}

\title{Deterministic spin-wave interferometer based on Rydberg blockade}
\author{Ran Wei}
\affiliation{Hefei National Laboratory for Physical Sciences at Microscale and Department
of Modern Physics, University of Science and Technology of China, Hefei,
Anhui 230026, China}
\author{Bo Zhao}
\email{bo.zhao@uibk.ac.at}
\affiliation{Institute for Theoretical physics, University of Innsbruck, A-6020
Innsbruck, Austria}
\affiliation{Institute for Quantum Optics and Quantum Information of the Austrian Academy
of Science, \\
A-6020 Innsbruck, Austria}
\author{Youjin Deng}
\email{yjdeng@ustc.edu.cn}
\affiliation{Hefei National Laboratory for Physical Sciences at Microscale and Department
of Modern Physics, University of Science and Technology of China, Hefei,
Anhui 230026, China}
\author{Yu-Ao Chen}
\affiliation{Fakult\"at f\"ur Physik, Ludwig-Maximilian-Universit\"at, Schellingstrasse
4, 80798 M\"unchen, Germany}
\affiliation{Max-Planck-Institut f\"ur Quantenoptik, Hans-Kopfermann-Strasse 1, 85748
Garching, Germany }
\author{Jian-Wei Pan}
\affiliation{Hefei National Laboratory for Physical Sciences at Microscale and Department
of Modern Physics, University of Science and Technology of China, Hefei,
Anhui 230026, China}

\begin{abstract}
The spin-wave (SW) NOON state is an $N$-particle Fock state with two atomic
spin-wave modes maximally entangled. Attributed to the property that the
phase is sensitive to collective atomic motion, the SW NOON state can be
utilized as a novel atomic interferometer and has promising application in
quantum enhanced measurement. In this paper we propose an efficient protocol
to deterministically produce the atomic SW NOON state by employing Rydberg blockade.
Possible errors in practical manipulations are
analyzed. A feasible experimental scheme is suggested. Our scheme is far
more efficient than the recent experimentally demonstrated one, which only
creates a heralded second-order SW NOON state.
\end{abstract}

\maketitle

\section{Introduction}

The NOON state, an $N$-particle Fock state with two modes maximally
entangled, has attracted many interests since it has the potential to
enhance the measurement precision by employing quantum entanglement \cite{lee2002}.
Attributed to the property of superresolution and supersensitivity,
the NOON state has been experimentally realized in various photonic systems
\cite{walther2004,mitchell2004,nagata2007,resch2007}. Recently, a new type of NOON
state - the atomic spin wave (SW) NOON state - was proposed, and a
heralded second-order SW NOON state as well, was experimentally demonstrated
\cite{yuao2010}. The scheme \cite{yuao2010} employs Raman transitions to
generate the atom-photon entanglement and the SW NOON state is created in a
herald way by detecting the photons. The SW NOON state can
be used as an atomic SW interferometer and can in principle be implemented
in a scalable way. However, owing to the probabilistic nature,
this SW interferometer works in a very low efficiency and thus
cannot be put into practical measurement.

In recent years, the Ryberg atom draw extensive concern in quantum
information processing \cite{molmer2010}. It has large size and can exhibit large
electric dipole moment. This property introduces strong interactions between
two Rydberg atoms. Consequently, in a small volume, when an atom is excited
to the Rydberg state $|r\rangle $, the energy level of state $|r\rangle $
for other atoms will be shifted by $\Delta_{e}$. Therefore, the probability for other
atoms being excited to $|r\rangle $ is suppressed by a factor of
$1/\Delta _{e}^{2}$. In the limit $\Delta_{e}\rightarrow\infty$,
only one atom is excited to $|r\rangle $. This is the so-called
Rydberg blockade mechanism. The Rydberg blockade has been proposed to
deterministically implement quantum computer and quantum repeater \cite
{jaksch2000,lukin2001,saffman2005a,saffman2005b,
saffman2009,markus2009,zhaobo2010,yanghan2010,isenhower2010}.

In this paper, we propose an efficient way to implement the SW
interferometer by deterministically generating the SW NOON state with Rydberg
blockade. An elaborate error analysis shows that the $20$th-order SW NOON state
can be generated with $91\%$ fidelity under realistic parameters,
and accordingly a high fidelity SW interferometer with $F\approx82\%$ can be realized.
This Rydberg-based SW interferometer is much more efficient than the one based on photon
detection and might be used as an inertial sensor, for measuring position and displacement,
or further, for measuring acceleration and platform rotation.
The remaining of this paper is organized as follows. Sec.~II describes an envisioned setup
and presents the scheme to generate and measure the SW NOON state. Error
analysis in practical implementations is given in Sec.~III. Experimental
realization is suggested in Sec.~IV, and finally we conclude in Sec.~V.

\section{Protocol}

\begin{figure}[tbh]
\includegraphics[width=9cm]{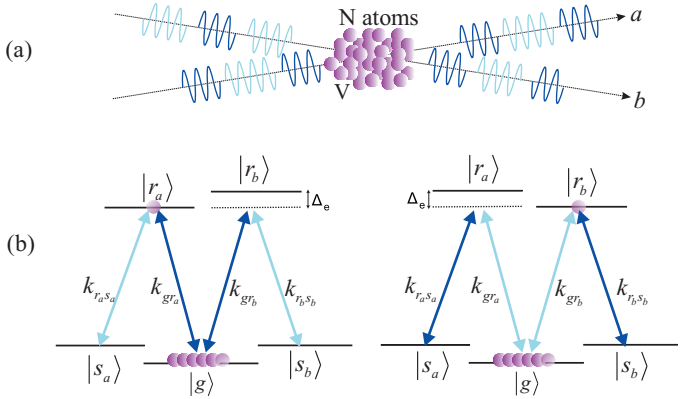}
\caption{(Color Online) $\bm{(a)}$ An ensemble of N atoms trapped in volume V.
The atoms are coupled by four types of light pulses,
propagating along two spatial modes $a,b$.
$\bm{(b)}$ The double-$\Lambda$ type energy levels. An effective energy
shift $\Delta _{e}$ is introduced because of the strong interaction between the
atoms at the Rydberg states.}
\label{figure1}
\end{figure}

We envision a setup as illustrated in Fig.~\ref{figure1}(a). An ensemble of $%
N$ atoms is confined in a volume $V$, where the blockade mechanism is effective.
In other words, the scale of $V$ is smaller than the blockade radius.
The working atomic energy levels are chosen to be of the double-$\Lambda$
type, as shown in Fig.~\ref{figure1}(b). They are labeled as the
ground state $|g\rangle $, the Rydberg state $|r_{a}\rangle$,
$|r_{b}\rangle $, and the metastable state $|s_{a}\rangle $, $|s_{b}\rangle$.
The atoms are coupled by four types of classic light pulses propagating along two
spatial modes $a,b$, whose wave vectors are denoted as $\bm{k}_{gr_{a}}$,
$\bm{k}_{r_{a}s_{a}}$, $\bm{k}_{gr_{b}}$ and $\bm{k}_{r_{b}s_{b}}$
respectively. They will also be used to denote the corresponding
light pulses if no ambiguity arises. These light pulses couple 
$|g\rangle$ and $|r_{a}\rangle$, $|r_{a}\rangle$ and $|s_{a}\rangle$, $|g\rangle$
and $|r_{b}\rangle$, and $|r_{b}\rangle$ and $|s_{b}\rangle$ respectively,
as illustrated in Fig.~\ref{figure1}(b).

Before giving the detailed scheme, we shall first introduce some definitions.
We define a collective ground state $|\bm{0}\rangle\equiv|g...g\rangle$,
a collective operator $M_{\bm{k},\epsilon}^{\dagger}\equiv\frac{1}{\sqrt{N}}
\sum\limits_{j}^{N}e^{i\bm{k}\cdot \bm{r}_{j}}|\epsilon_{j}\rangle
\langle g|$, and $|\bm{1},\bm{k}\rangle _{\epsilon }$ $(\epsilon
=r_{a},r_{b},s_{a},s_{b})$ to describe a collective state with wave vector
$\bm{k}$,
\begin{equation}
|\bm{1},\bm{k}\rangle _{\epsilon }\equiv \frac{1}{\sqrt{N}}
\sum\limits_{j}^{N}e^{i\bm{k}\cdot \bm{r}_{j}}|g...\epsilon
_{j}...g\rangle =M_{\bm{k},\epsilon }^{\dagger }|\bm{0}\rangle.
\end{equation}
Namely, state $|\bm{1},\bm{k}\rangle _{\epsilon }$ is a coherent
superposition of states which have a specific atom at $|\epsilon \rangle $
with the position-dependent phase under the wave vector $\bm{k}$. The same
applies to the higher-order collective state $|\bm{\ell},\bm{k}\rangle _{\epsilon
}\equiv \frac{1}{\sqrt{\ell!}}(M_{\bm{k},\epsilon }^{\dagger })^{\ell
}|0\rangle_{\epsilon}$, with $\ell$ a positive integer. On this basis,
a $\ell$th-order SW NOON state can be written as
\begin{equation}
\left\vert \mathrm{NOON}\right\rangle _{\ell }=\frac{1}{\sqrt{2}}\left(
\left\vert \bm{\ell} ,\bm{k}\right\rangle _{s_{a}}+\left\vert\bm{\ell} ,%
\bm{k}\right\rangle _{s_{b}}\right).  \label{eq_SW}
\end{equation}

We first consider the ideal case by making the following assumptions.
(1), the atom number is exactly known, i.e., $\Delta N=0$; (2), the Rydberg
blockade mechanism is perfect, i.e., $\Delta _{e}\rightarrow \infty $; (3),
the lifetime of the Rydberg state is infinite and thus no spontaneous decay
occurs; (4), the atomic cloud remains still during the whole process. On
this basis, our scheme to generate a $\ell $th-order SW NOON state can be
described as

\begin{enumerate}
\item Prepare an ensemble at the ground state $|\bm{0}\rangle $.

\item Apply sequentially a collective $\pi$ pulse $\bm{k}_{gr_a}$
and a single-atomic $\pi/2$ pulse $\bm{k}_{r_as_a}$. The former flips
one of the $N$ atoms from $|\bm{0}\rangle$ to the Rydberg state
$|\bm{1},\bm{k}_{gr_{a}}\rangle_{r_a}$
and the latter flips $|\bm{1},\bm{k}_{gr_{a}}\rangle_{r_a}$ to
the equal superposition of the first-order SW state
$|\bm{1},\bm{k}_{gr_as_a}\rangle_{s_a}$
and $|\bm{1},\bm{k}_{gr_{a}}\rangle_{r_a}$, where 
$\bm{k}_{\epsilon_1\epsilon_2\epsilon_2}\equiv
\bm{k}_{\epsilon_1\epsilon_2}-\bm{k}_{\epsilon_2\epsilon_3}$ 
$(\epsilon_1,\epsilon_2,\epsilon_3=g,r_{a},r_{b},s_{a},s_{b})$. 
Accordingly, one obtains
\begin{equation*}
i|\bm{1},\bm{k}_{gr_as_a}\rangle _{s_{a}}+|\bm{1},
\bm{k}_{gr_{a}}\rangle _{r_{a}},
\end{equation*}
where a relative phase shift $\pi/2$ is introduced.

\item Apply successively three collective $\pi$ pulses
$\bm{k}_{gr_b}$, $\bm{k}_{gr_a}$ and $\bm{k}_{gr_b}$, which leads to
\begin{eqnarray*}
&&|\bm{1},\bm{k}_{gr_{a}s_{a}}\rangle _{s_{a}}|\bm{1},\bm{k}%
_{gr_{b}}\rangle _{r_{b}}-|\bm{1},\bm{k}_{gr_{a}}\rangle _{r_{a}}  \\
&&\qquad \qquad \qquad \bm{\Downarrow}   \\
&&i|\bm{1},\bm{k}_{gr_{a}s_{a}}\rangle _{s_{a}}|\bm{1},\bm{k}%
_{gr_{b}}\rangle _{r_{b}}+|\bm{0}\rangle   \\
&&\qquad \qquad \qquad \bm{\Downarrow}   \\
&&i|\bm{1},\bm{k}_{gr_{a}s_{a}}\rangle _{s_{a}}+
|\bm{1},\bm{k}_{gr_{b}}\rangle _{r_{b}}.
\end{eqnarray*}

\item Apply in order a collective $\pi$ pulse $\bm{k}_{gr_a}$
and a single-atomic $\pi$ pulse $\bm{k}_{r_bs_b}$, and a collective
$\pi$ pulse $\bm{k}_{gr_b}$ and a single-atomic $\pi$ pulse $\bm{k}_{r_as_a}$,
which results in
\begin{eqnarray*}
&&|\bm{1},\bm{k}_{gr_{a}s_{a}}\rangle _{s_{a}}
|\bm{1},\bm{k}_{gr_{b}}\rangle _{r_{a}}-
|\bm{1},\bm{k}_{gr_{b}}\rangle_{r_{b}}   \\
&&\qquad \qquad \qquad \qquad \bm{\Downarrow}   \\
&&|\bm{1},\bm{k}_{gr_{a}s_{a}}\rangle _{s_{a}}
|\bm{1},\bm{k}_{gr_{b}}\rangle _{r_{a}}-
i|\bm{1},\bm{k}_{gr_{b}s_{b}}\rangle_{s_{b}}\\
&&\qquad \qquad \qquad \qquad \bm{\Downarrow}   \\
&&|\bm{1},\bm{k}_{gr_{a}s_{a}}\rangle _{s_{a}}
|\bm{1},\bm{k}_{gr_{b}}\rangle _{r_{a}}+
|\bm{1},\bm{k}_{gr_{b}s_{b}}\rangle
_{s_{b}}|\bm{1},\bm{k}_{gr_{b}}\rangle _{r_{b}}  \\
&&\qquad \qquad \qquad \qquad \bm{\Downarrow}   \notag \\
&&i|\bm{2},\bm{k}_{gr_{a}s_{a}}\rangle _{s_{a}}+|\bm{1},
\bm{k}_{gr_{b}s_{b}}\rangle_{s_{b}}|\bm{1},\bm{k}_{gr_{b}}\rangle _{r_{b}}.
\end{eqnarray*}

\item Repeatedly apply a sequence of four collective $\pi$ pulses $\bm{k}_{gr_a}$,
$\bm{k}_{r_bs_b}$, $\bm{k}_{gr_b}$, $\bm{k}_{r_as_a}$ for $\ell-2$ times, and one obtains
\begin{eqnarray*}
&&|\bm{\ell},\bm{k}_{gr_{a}s_{a}}\rangle_{s_{a}}+
|\bm{\bm{\ell}-1},\bm{k}_{gr_{b}s_{b}}\rangle
_{s_{b}}|\bm{1},\bm{k}_{r_{b}s_{b}}\rangle _{r_{b}}.
\end{eqnarray*}

\item Apply a collective $\pi $ pulse to flip the
atom from $|\bm{\bm{\ell}-1},\bm{k}_{gr_{b}s_{b}}\rangle_{s_{b}}
|\bm{1},\bm{k}_{r_{b}s_{b}}\rangle _{r_{b}}$ to
$|\bm{\ell},\bm{k}_{gr_{b}s_{b}}\rangle _{s_{b}}$
and take into account the normalized factor,
and one obtains a $\ell$th-order SW NOON state
\begin{equation}
|\Psi\rangle_{\ell}=(|\bm{\ell},\bm{k}_{gr_{a}s_{a}}\rangle_{s_{a}}+
|\bm{\ell},\bm{k}_{gr_{b}s_{b}}\rangle _{s_{b}})/\sqrt{2}.
\end{equation}
\end{enumerate}

According to the above procedure, the generation of a $\ell$th-order SW NOON
state needs totally $4\ell +2$ light pulses, the number of which is linear to $\ell $.
Note that one needs two $\pi $ pulses $\bm{k}_{gr_a}$ and $\bm{k}_{r_as_a}$ to
produce a first-order SW state $|\bm{1},\bm{k}_{gr_{a}s_{a}}\rangle_{s_{a}}$. Accordingly,
two $\ell $th-order SW states $|\bm{\ell},\bm{k}_{gr_{a}s_{a}}\rangle_{s_{a}}$
and $|\bm{\ell},\bm{k}_{gr_{b}s_{b}}\rangle_{s_{b}}$ would consume
$4\ell$ light pulses. The $\ell$th-order SW NOON state is the superposition of
two $\ell$th-order SW states at the $a$ and $b$ modes. Thus, we consider the
above protocol close to being optimal, albeit the possibility of further
improvement is not entirely excluded.

Here we demonstrate how the SW NOON state can be utilized as an atomic
interferometer. Let's assume that, after the $\ell$th-order SW NOON state is
prepared, the atomic cloud moves to a new position with a displacement $\Delta\bm{x}$.
To measure $\Delta\bm{x}$, we apply a sequence of operations reverse to the
generation procedure, until the last operation, i.e., the collective $\pi$
pulse $\bm{k}_{gr_a}$. Detailed calculations show that we obtain the
superposition state
\begin{eqnarray}
\label{eq_final}
\notag|\Psi'\rangle_{\ell}&=&(ie^{i(\bm{k}_{gr_as_a})
\cdot\Delta\bm{x}}(1+e^{i\ell\Delta\bm{k}\cdot\Delta\bm{x}})
|\bm{1},\bm{k}_{gr_as_a}\rangle_{s_a}\\
&+&e^{i\bm{k}_{gr_b}\cdot\Delta\bm{x}}(1-e^{i\ell\Delta\bm{k}
\cdot\Delta\bm{x}})|\bm{1},\bm{k}_{gr_a}\rangle_{r_a})/\sqrt{2},
\end{eqnarray}
where $\Delta\bm{k}\equiv-\bm{k}_{gr_as_a}-\bm{k} _{gr_bs_b}$.
Note that, by applying an ionizing electric field, the
Rydberg state $|\bm{1},\bm{k}_{gr_a}\rangle_{r_a}$ will be ionized and a free
electron will fly out of the atomic ensemble. Thus, the state~(\ref{eq_final})
can be measured onto the $|\bm{1},\bm{k}_{gr_a}\rangle_{r_a}$ basis, and the
average result will reflect the phase shift $\ell\Delta\bm{k}\cdot\Delta\bm{x}$.
Since the wave vectors of the light pulses are known, this gives
the displacement $\Delta \bm{x}$. The phase shift is proportional to the
order $\ell$, and thus the larger $\ell$ would bring $\Delta\bm{x}$ the
better precision.

\section{Error analysis}

In actual implementations, errors can always occur. For instance, the precise
number $N$ of atoms in the ensemble is normally unknown, and the atom number
$N$ also varies for different experimental trials. This leads to an
uncertainty $\Delta N$ of the atom number, which is $\Delta N\simeq\sqrt{N}$
for large $N$. Since the collective Rabi frequency $\Omega_c$ of
the $\pi$ pulse $\bm{k}_{gr_{\lambda}}$ \cite{definition}
is related to the atom number $N$ as $\Omega_c\propto\sqrt{N}$,
$\Delta N$ would induce an imprecision in $\Omega_c$ as $\Delta\Omega_c/
\Omega_c\simeq1/(2\sqrt{N})$.
This means that, when a collective $\pi$ pulse $\bm{k}_{gr_{\lambda}}$
is applied to flip one of the atoms from $|\bm{0}\rangle $ to
$|\bm{1},\bm{k}_{gr_{\lambda}}\rangle_{r_{\lambda}}$, there
exists a probability $p\simeq\pi^2/(16N)$ that the flip fails. To generate a
$\ell$th-order SW NOON state, the total error introduced by
$\Delta N$ is about $\pi^2\ell/(8N)$. In lab, one
can prepare an ensemble of $N\approx400$ atoms, and thus the error is
about $\pi^2\ell/(8N)\approx6\%$ for order $\ell=20$.

Aside from the error induced by the uncertainty of the atom number,
the imperfect blockade mechanism and the finite lifetime of the Rydberg state
also introduces errors. Attributed to these factors,
each operation in our scheme is implemented with a non-unity probability.
We step by step analyze all the operations from Step $1$ to Step $6$,
and find that, these non-unity probabilities can be categorized
into five types, denoted as $P^{I}$, $P^{II}$, $P^{III} $, $P^{IV}_q$, $P^{V}_q$,
and the generated $\ell$th-order SW NOON state should be rewritten approximately as
\begin{align}
\sqrt{\mathcal{P}_{\ell}(P^{I}P^{II}P^{III})^{\ell}}|\Psi\rangle_{\ell},
\label{eq_p1}
\end{align}
where $\mathcal{P}_{\ell}=\prod_{q=1}^{\ell}P^{IV}_qP^{V}_q$.
Symbol $q$ stands for the order of the SW state during the generation process,
and it increases from $1$ to $\ell$ as one produces the $\ell$th-order SW NOON state.
Accordingly, the probability for preparing the $\ell$th-order SW NOON state is
\begin{align}
P(\ell)=\mathcal{P}_{\ell}(P^{I}P^{II}P^{III})^{\ell}.
\label{eq_p2}
\end{align}
The total error accumulated by these operations is the probability that one fails to 
generate the $\ell$th-order SW NOON state, thus it reads $E(\ell)=1-P(\ell)$.
(The error induced by the uncertainty of atom number is not included in $E(\ell)$.)
Before evaluating $E(\ell)$, we shall first analyze the origins of these probabilities.

The probability $P^{I}$ is introduced by the imperfect blockade that occurs
between the atoms of the same mode when the pulse $\bm{k}_{gr_{\lambda}}$ flips
one of the atoms from $|\bm{0}\rangle$ to $|\bm{1},\bm{k}_{gr_{\lambda}}\rangle_{r_{\lambda}}$.
In other words, there is an error that two atoms are excited to the Rydberg
state $|\bm{2},\bm{k}_{gr_\lambda}\rangle_{r_{\lambda}}$ due to the non-infinite energy
shift. This mechanics is described by the following equations,
\begin{align}
\label{e1}
i\dot{c}_0 &=-\frac{\sqrt{N}\Omega}{2}c_1, \\
\label{e2}
i\dot{c}_1 &=-i\frac{\gamma}{2}c_1-\frac{\sqrt{N}\Omega}{2}c_0
-\frac{\sqrt{2N}\Omega}{2}c_2, \\
\label{e3}
i\dot{c}_2 &=(\Delta_e-i\gamma)c_2-\frac{\sqrt{2N}\Omega}{2} c_1,
\end{align}
where $c_0,c_1,c_2$ stand for the amplitudes of $|\bm{0}\rangle$,
$|\bm{1},\bm{k}_{gr_\lambda}\rangle_{r_{\lambda}}$, $|\bm{2},\bm{k}_{gr_\lambda}\rangle_{r_{\lambda}}$.
Symbols $\gamma/2$ and $\gamma$ are the decay rates of
$|\bm{1},\bm{k}_{gr_\lambda}\rangle_{r_{\lambda}}$ and $|\bm{2},\bm{k}_{gr_\lambda}\rangle_{r_{\lambda}}$.
Symbol $\Delta_e$ is the effective finite energy shift, and $\sqrt{N}\Omega,\sqrt{2N}\Omega$ are the
corresponding two collective Rabi frequencies, which have been assumed to be real.
Since the amplitudes for the states of more than two atoms being excited are
significantly suppressed due to the Rydberg blockade, we have neglected them
here and in the following. Besides, we have assumed the number of
atoms $N\gg1$ and the coupling light pulses are all in resonance. The initial
condition describing this mechanics is $c_0(0)=1,c_1(0)=0,c_2(0)=0$. After
applying the collective $\pi$ pulse $\bm{k}_{gr_{\lambda}}$ with
the operation time $\Delta t=\pi/(\sqrt{N}\Omega)$, one can express the
probability for generating $|\bm{1},\bm{k}_{gr_\lambda}\rangle_{r_{\lambda}}$
from $|\bm{0}\rangle$, as $P^{I}=|c_1(\Delta t)|^2$.

The probability $P^{II}$ characterizes the imperfect blockade that takes
place between the atoms of the different modes during $\Delta t$. That is to say,
there is an error that the pulse $\bm{k}_{gr_{\lambda}}$
would flip one of the atoms from $|\bm{0}\rangle$ to $|\bm{1},\bm{k}_{gr_\lambda}\rangle_{r_{\lambda}}$
when another atom has already been excited to $|\bm{1},\bm{k}_{gr_{\bar{\lambda}}}\rangle_{r_{\bar{\lambda}}}$.
Accordingly, this mechanics is governed by the following equations,
\begin{align}
\label{e4}
i\dot{c}_0 &=-\frac{\sqrt{N}\Omega}{2}c_1, \\
\label{e5}
i\dot{c}_1 &=(\Delta_e-i\frac{\gamma}{2})c_1-\frac{\sqrt{N}\Omega}{2} c_0.
\end{align}
The initial condition describing this mechanics is $c_0(0)=1,c_1(0)=0$,
and one can express the probability for holding the atoms
at the ground state, as $P^{II}=|c_0(\Delta t)|^2$.

The probability $P^{III}$ is contributed by the decay rate of the Rydberg
state. The finite lifetime will inevitably cause some loss when the atom is
still at the Rydberg state during $\Delta t$, thus the probability for the
atom remaining at the Rydberg state is $P^{III}=e^{-\gamma\Delta t}$.

These three types ($P^{I}$, $P^{II}$, $P^{III}$) are all determined by
a shared Rabi frequency $\Omega$ or a shared operation time $\Delta t$. Note
that there is tradeoff between the imperfect Rydberg blockade and the loss
caused by the decay, and a simple argument is that if we enhance the the
magnitude of the Rabi frequency to shorten the operation time,
which reduces the loss from the Rydberg state, it will be associated with
more errors from the imperfect blockade. Therefore, there is an optimal Rabi
frequency to maximize the value of $P^{I}P^{II}P^{III}$.
By numerically solving Eqs.~(\ref{e1}-\ref{e3}) and Eqs.~(\ref{e4}-\ref{e5}),
one can easily obtain this maximal value.

The probability $P^{IV}_q$ reflects an error that one of the atoms at $|\bm{q-1},\bm{k}_{gr_{\lambda}s_{\lambda}}\rangle_{s_{\lambda}}
|\bm{1},\bm{k}_{gr_{\lambda}}\rangle_{r_{\lambda}}$
would be flipped back to $|\bm{q-2},\bm{k}_{gr_{\lambda}s_{\lambda}}
\rangle_{s_{\lambda}}|\bm{2},\bm{k}_{gr_{\lambda}}\rangle_{r_{\lambda}}$
when the pulse $\bm{k}_{r_{\lambda}s_{\lambda}}$ is applied to
flip the atom from $|\bm{q-1},\bm{k}_{gr_{\lambda}s_{\lambda}}
\rangle_{s_{\lambda}}|\bm{1},\bm{k}_{gr_{\lambda}}\rangle_{r_{\lambda}}$
to $|\bm{q},\bm{k}_{gr_{\lambda}s_{\lambda}}\rangle_{s_{\lambda}}$.
This mechanics is described by the following equations,
\begin{align}
\label{e6}
i\dot{\widetilde{c}}_0 &=-\frac{\sqrt{q}\widetilde{\Omega}}{2}\widetilde{c}_1,\\
\label{e7}
i\dot{\widetilde{c}}_1 &=-i\frac{\gamma}{2}\widetilde{c}_1-\frac{\sqrt{q}
\widetilde{\Omega}}{2} \widetilde{c}_0-\frac{\sqrt{2(q-1)}\widetilde{\Omega}
}{2}\widetilde{c}_2,\\
\label{e8}
i\dot{\widetilde{c}}_2 &=(\Delta_e-i\gamma)\widetilde{c}_2-
\frac{\sqrt{2(q-1)}\widetilde{\Omega}}{2}\widetilde{c}_1,
\end{align}
where $\widetilde{c}_0,\widetilde{c}_1,\widetilde{c}_2$ are the amplitudes
of $|\bm{q},\bm{k}_{gr_{\lambda}s_{\lambda}}\rangle_{s_{\lambda}}$,
$|\bm{q-1},\bm{k}_{gr_{\lambda}s_{\lambda}}\rangle_{s_{\lambda}}
|\bm{1},\bm{k}_{gr_{\lambda}}\rangle_{r_{\lambda}}$,
$|\bm{q-2},\bm{k}_{gr_{\lambda}s_{\lambda}}\rangle_{s_{\lambda}}
|\bm{2},\bm{k}_{gr_{\lambda}}\rangle_{r_{\lambda}}$.
Symbols $\sqrt{q}\widetilde{\Omega},\sqrt{2(q-1)}\widetilde{\Omega}$ are the
corresponding two collective Rabi frequencies, which have also been assumed
to be real. The initial condition describing this mechanics is
$\widetilde{c}_0(0)=0,\widetilde{c}_1(0)=1,\widetilde{c}_2(0)=0$. After applying the
collective $\pi$ pulse $\bm{k}_{r_{\lambda}s_{\lambda}}$ with the operation time
$\Delta\widetilde{t}_q=\pi/(\sqrt{q}\widetilde{\Omega})$,
one can express the probability for producing the $q$th-order SW state $|\bm{q},\bm{k}_{gr_{\lambda}s_{\lambda}}\rangle_{s_{\lambda}}$,
as $P^{IV}_q=|\widetilde{c}_0(\Delta\widetilde{t}_q)|^2$.

The origin of $P^{V}_q$ is similar to $P^{III}$,
it reflects the probability that the atom remains at
the Rydberg state during $\Delta\widetilde{t}_q$, and thus
$P^{V}_q=e^{-\gamma\Delta\widetilde{t}_q}$. The value of $P^{IV}_qP^{V}_q$ is
determined by a shared Rabi frequency $\widetilde{\Omega}$ or a shared operation time
$\Delta\widetilde{t}_q$. Likewise, one can calculate the maximal value of $P^{IV}_qP^{V}_q$
by numerically solving Eqs.~(\ref{e6}-\ref{e8}) with $q$ from $1$ to $\ell$.

\begin{figure}[!hbt]
\includegraphics[width=8.5cm]{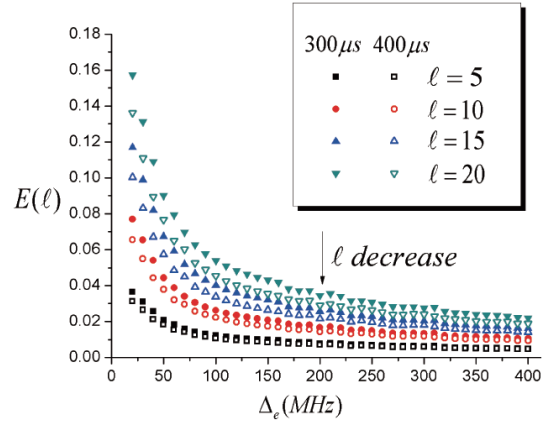}
\caption{(Color Online) The figure demonstrates the error $E(\ell)$ versus
the energy shift $\Delta_e$ under various order $\ell$ after generating the
SW NOON state. The solid data and the open one respectively denote the
lifetime of the Rydberg state with $\protect\tau=300$ $\protect\mu s$ and
$\protect\tau=400$ $\protect\mu s$.}
\label{figure2}
\end{figure}

To evaluate $E(\ell)$, we choose the parameters as, the atom number $N=400$, the lifetime of
the Rydberg state $\tau=1/(2\pi\gamma)=300$ $\mu s$ and $400$ $\mu s$,
and the energy shift $\Delta_e$ varying from $20$ $MHz$ to
$400$ $MHz$. Accordingly, Eq.(\ref{eq_p2}) can be calculated in a numerical way.
We obtain the error $E(\ell)$ versus the energy shift $\Delta_e$, shown in Fig.\ref{figure2}.

From the figure, we see that the larger the energy shift, the smaller the error, and the error vanishes as $\Delta_e$ tends to infinity. This is an anticipated result since the error $E\sim\Omega^2/\Delta_e^2$. However, in actual experiment, $\Delta_e$ cannot be unlimitedly large. An intrinsic limitation originates from the average distance of two Rydberg atoms, which should be larger than the radius of each Rydberg atom.  In the limit of high density where the Rydberg atoms remarkably overlap, our blockade model is inappropriate, and a more elaborate mechanism should be taken into account. This mechanism goes beyond the extent of our paper and will not be discussed. Besides, as one readily expects, the figure shows that the error is suppressed as the lifetime of the Rydberg state becomes longer, and is intensified when the order $\ell$ of the SW NOON state increases.

\section{Experimental realization}

To design an atomic interferometer with sufficiently high precision and
relatively high fidelity, we use the $20$th-order SW NOON state for the
practical application. The interferometer can be implemented by cold
alkali atoms. By choosing the suitable laser polarization, the two spacial
modes $a$ and $b$ can be individually addressed. 
The energy shift is isotropic due to the property of repulsive van der Waals interaction. 
The lifetime of the Rydberg state with $\tau=300\sim400$ $\mu s$ is achievable
by exciting the atoms to the Rydberg $s$ state with a principal quantum
number $n=100$~\cite{saffman2005a}. In our scheme, the energy shift
$\Delta_e $ of the Rydberg state can be expressed as
$\Delta_e=-n^{11}(c_0+c_1n+c_2n^2)/r^6 $~\cite{singer2005}, where the terms
$1/r^8$ and $1/r^{10}$ are neglected due to the dominating long-range
property. For Rubidium, $c_0=13,c_1=-0.85,c_2=0.0034$~\cite{singer2005}, and
thus an ensemble of atoms with the radius $R=3.8$ $\mu m$ enables the energy
shift $\Delta_e\geq300$ $MHz$, which ensures the error $E(20)<3\%$, as illustrated in
Fig.\ref{figure2}. In a volume of $4\pi/3R^3$, a density of $1.7\times10^{12}$
$cm^{-3}$ allows $N\approx400$ atoms in an ensemble. Based on these
estimated parameters above, we suggest to employ the one-dimensional
optical lattice as the experimental setup, where the size of the ensemble
can be controlled by tuning the angle between the trapping light fields
\cite{fallani2005}. Finally, we should point out that, to detect the
displacement of atomic cloud by the interferometer, the reverse operations to those in the
generation procedure should be considered, and thus the total error is
doubled. Fortunately, the field ionization can be implemented with near-unity
detection efficiency \cite{guerlin2007}. Therefore, taking into account
the error induced by the uncertainty of atom number, our proposed atomic SW interferometer
with a high precision ($\ell=20$) can reach a high fidelity as $F\approx1-2\times(6\%+3\%)=82\%$.

\section{Summary}

By employing Rydberg blockade, we have demonstrated an efficient scheme to
deterministically produce the atomic SW NOON state, of which, a direct
application is the atomic SW interferometer. Possible errors in practical
manipulations are analyzed, and the experimental realization also is
suggested. Our proposed atomic SW interferometer is far more efficient than the
recent experimentally demonstrated one, and holds promise in the
practical application.

\section{Acknowledgement}
This work is supported by the NNSFC, the NNSFC of Anhui (under Grant No.
090416224), the CAS, the National Fundamental Research Program (under Grant
No. 2011CB921304), and the SFB FOQUS of FWF.


\begin{thebibliography}{99}
\bibitem{definition} For the compactness, 
we define the notation $\{\lambda,\bar{\lambda}\}\equiv\{a,b\},\{b,a\}$.
\end{thebibliography}
\end{document}